\documentclass[final,1p,times]{elsarticle}
\journal{Mediterranean Journal of Mathematics}
\usepackage{tikz-cd}
\usepackage{extpfeil}
\usepackage{amssymb}
\usepackage{amsthm}
\usepackage{latexsym}
\usepackage{amsmath}
\usepackage{color}
\usepackage{graphicx}
\usepackage{indentfirst}
\usepackage{mathrsfs}
\usepackage{lipsum}
\allowdisplaybreaks

\newtheorem{theorem}{\color{black}\indent \textbf{Theorem}}[section]

\newtheorem{proposition}{\color{black}\indent Proposition}[section]
\newtheorem{definition}{\color{black}\indent Definition}[section]
\newtheorem{remark}{\color{black}\indent Remark}[section]
\newtheorem{corollary}{\color{black}\indent Corollary}[section]
\newtheorem{example}{\color{black}\indent Example}[section]

\begin{document}
	
	\begin{frontmatter}

		\title{Noether-Type Theorems and the Generalized Herglotz Principle in q-Contact Geometry}
	\author{
Melvin Leok$^{a}$,
Cristina Sardón$^{b}$,
Xuefeng Zhao$^{c}$ \\[1ex]
$^{a}$Department of Mathematics, University of California, San Diego, 
9500 Gilman Drive, Dept. 0112, La Jolla, CA 92093-0112, USA \\
$^{b}$Department of Applied Mathematics, Universidad Politécnica de Madrid, 
Av. Juan de Herrera 6, 28040, Madrid, Spain \\
$^{c}$College of Mathematics, Jilin University, Changchun, 130012, P. R. China \\[1ex]
\texttt{mleok@ucsd.edu}, \texttt{mariacristina.sardon@upm.es}, \texttt{zhaoxuef@jlu.edu.cn}
}
	\begin{abstract}
We develop a unified geometric framework for dissipative mechanical systems based on uniform $q$-contact manifolds, which provide an extended phase space equipped with multiple contact $1$-forms. Within this setting, we construct both Hamiltonian and Lagrangian formalisms and establish a generalized Noether-type theorem describing the relationship between symmetries and dissipated quantities. 

We further show that $q$-contact Lagrangian systems admit a genuine variational origin through a generalized Herglotz principle involving multiple action variables. The resulting $q$-contact Euler--Lagrange equations naturally depend on the scalar combination $\sum_{i=1}^q \partial L/\partial z_i$, reflecting the intrinsic structure of uniform $q$-contact geometry. We prove that this variational formulation is fully equivalent to the geometric $q$-contact Hamiltonian dynamics generated by the energy function.

Several explicit examples involving multi-parameter dependent dynamics illustrate the effectiveness of the theory and demonstrate its potential to provide geometric insight into complex dissipative systems, thereby extending the scope of classical Lagrangian mechanics beyond symplectic and single-contact structures.
\end{abstract}
		
		\begin{keyword}
			$q$-Contact  manifold, Symmetry, Hamiltonian system, $q$-Lagrangian system, dissipated quantity.
		\end{keyword}
	\end{frontmatter}

\section{Introduction}

We consider a unified framework based on $q$-contact manifolds, which serve as an extended phase space incorporating multiple contact $1$-forms. Within this setting, we develop both Hamiltonian and Lagrangian formalisms, establish a generalized version of Noether's theorem that captures the interplay between symmetries and dissipative behavior, and provide a variational foundation for the theory through a $q$-contact extension of the Herglotz principle. To demonstrate the applicability of the framework, we construct explicit examples involving multi-parameter dependent dynamics. These examples validate the proposed theory and highlight its potential in providing geometric insight into complex dissipative systems, thereby extending the reach of traditional Lagrangian mechanics on manifolds with contact-type structures.

Physical systems inevitably experience external forces, energy dissipation, damping, and various irreversible effects observed in real-world scenarios. Developing appropriate models for dissipative systems is therefore essential, as such features play a crucial role in shaping the dynamics of physical systems across numerous fields, including quantum mechanics \cite{Hooft} and general relativity \cite{Cariglia}.

The Noether theorem establishes a fundamental link between the symmetries of a Lagrangian system and the conserved quantities of the corresponding Euler–Lagrange equations. In its simplest form, the presence of a cyclic coordinate implies the conservation of the corresponding momentum. Specifically, if the Lagrangian $L = L(q^i, \dot{q}^i)$ is independent of a particular coordinate $q^j$, then the Euler–Lagrange equation
\begin{align}
	\frac{d}{dt} \left( \frac{\partial L}{\partial \dot{q}^j} \right) - \frac{\partial L}{\partial q^j} = 0,
\end{align}
implies that (see \cite{Arnold})
\begin{align}
	\dot{p}_j = \frac{d}{dt} \left( \frac{\partial L}{\partial \dot{q}^j} \right) = 0.
\end{align}

The Noether theorem can also be formulated geometrically \cite{Aldaya,Aldaya2,Cantrijn,Carinena,Carinena2,de2,de3,Prince1,Prince2,Sarlet,Zhao5}. In this setting, the Lagrangian $L$ is viewed as a function on the tangent bundle $TQ$ of the configuration manifold $Q$, and $X$ denotes a vector field on $Q$. Let $X^C$ and $X^V$ represent the complete and vertical lifts of $X$ to $TQ$, respectively. Then (see \cite{de4}):

\begin{theorem}[Noether]
	The condition $X^C(L) = 0$ holds if and only if $X^V(L)$ is a conserved quantity.
\end{theorem}

Here we adopt the symplectic formulation of Lagrangian mechanics. The Lagrangian $L$ induces a symplectic form on $TQ$ via
\begin{align}
	\omega_L = -d\alpha_L, \quad \alpha_L = S^*(dL),
\end{align}
where $L$ is assumed to be regular, $S$ is the canonical vertical endomorphism on $TQ$, and $S^*$ is its adjoint. The dynamics of the system are governed by
\begin{align}
	i_{\xi_L} \omega_L = dE_L,
\end{align}
where $E_L = \Delta(L) - L$ is the energy function associated with $L$, and $\Delta$ is the canonical Liouville vector field on $TQ$. The integral curves of the second-order differential equation $\xi_L$ project onto $Q$ as solutions of the Euler–Lagrange equations.

When dissipation is present, however, the symplectic framework is no longer sufficient. A natural variational extension of Hamilton’s principle capable of producing nonconservative dynamics was introduced by Herglotz. In the classical (single-contact) case, the Herglotz principle replaces the action integral by a dynamical variable $z(t)$ satisfying
\[
\dot z = L(q,\dot q,z), \qquad z(t_0)=z_0,
\]
and extremizes the terminal value $z(t_1)$. This yields the Herglotz–Euler–Lagrange equations
\[
\frac{d}{dt}\!\left(\frac{\partial L}{\partial \dot q^k}\right)
-
\frac{\partial L}{\partial q^k}
=
\frac{\partial L}{\partial z}
\frac{\partial L}{\partial \dot q^k},
\]
which coincide with the Lagrangian equations of contact mechanics. Thus, contact geometry admits a genuine variational origin \cite{herg1,herg2,herg3,herg4,herg5,herg6}.

As a higher-order generalization of contact geometry, a $q$-contact manifold is endowed with $q$ independent contact $1$-forms. This structure allows one to describe systems with multiple dissipation channels or constraints in a unified geometric manner. The notion of $q$-contact manifolds originated in Almeida’s doctoral dissertation \cite{Almeida}, motivated by the study of $\mathbb{R}^q$-Anosov actions and the Verjovsky conjecture (see \cite{Barbot}). Subsequently, Finamore established the Weinstein conjecture for uniform $q$-contact structures \cite{Finamore,Finamore2}, further demonstrating the relevance of this geometry.

In the present work, we extend the Herglotz variational principle to the $q$-contact setting by introducing $q$ action variables $z_i(t)$ satisfying
\[
\dot z_i = L(q,\dot q,z_1,\dots,z_q), \qquad i=1,\dots,q,
\]
and extremizing a terminal functional of the form $\sum_{i=1}^q z_i(t_1)$. The resulting $q$-contact Herglotz equations take the form
\[
\frac{d}{dt}
\left(
\frac{\partial L}{\partial \dot q^k}
\right)
-
\frac{\partial L}{\partial q^k}
=
\left(
\sum_{i=1}^q
\frac{\partial L}{\partial z_i}
\right)
\frac{\partial L}{\partial \dot q^k},
\]
and naturally involve the scalar combination $\sum_{i=1}^q \partial L/\partial z_i$, which is characteristic of uniform $q$-contact geometry. We prove that these equations coincide with the integral curves of the $q$-contact Hamiltonian vector field associated with the energy function $E_L$. In this way, the geometric and variational formulations are shown to be fully equivalent. 

The paper is organized as follows. In Section~2, we review the definition and basic properties of $q$-contact manifolds and define Hamiltonian systems on them. In Section~3, we develop a geometric framework for $q$-contact Lagrangian systems on the extended phase space $TQ \times \mathbb{R}^q$. In Section~4, we establish Noether-type results relating symmetries and dissipative quantities. In Section 5, we present the Herglotz variational principle on $q$-contact manifolds. In Section 6, we present an application: a controlled propulsion system with distributed dissipation.

	\section{ Hamiltonian systems on $q$-contact manifold}
	Firstly, we recall some results in contact geometry. Detailed proofs can be found in \cite{Perez,Willett}. Let $(M, \eta)$ be a contact manifold. This means that
	$M$ is a $(2n+1)$-dimensional manifold and $\eta\wedge (d\eta)^n$ is a volume form. Then, there exists a unique vector field $\mathcal R$ (the so-called Reeb vector field) such that
	$$i_{\mathcal R}d\eta=0,\quad i_{\mathcal R}\eta=1.$$
	Some authors \cite{Arnold,Le,Loose,Willett} define a contact manifold $(M,\mathcal H)$ as an odd-dimensional manifold $M$ and a contact distribution, that is, $\mathcal H$ is locally the kernel of a contact form $\eta$. Of course, every contact manifold $(M,\eta)$ is a contact manifold in this wider sense by taking $\mathcal H=\ker\eta.$  Conversely, a contact distribution $\mathcal H$ is globally the kernel of a contact form if and only if $\mathcal H$ is co-orientable.
	
A \emph{contact foliation} is a generalization of the flow of the Reeb vector field associated with a co-orientable contact structure. The structure that defines the contact foliation is known as a $q$-contact structure, which is defined as follows.
\begin{definition}(q-contact manifolds \cite{Almeida,Finamore,Finamore2})
	Let $n,q$ be positive integers and consider a $2n+q$ dimensional differential manifold $M$. A $q$-contact structure on $M$ is a collection $\vec{\lambda}=(\lambda_1,...,\lambda_q)$ of $q$ (pointwise) linearly independent non-vanishing $1$-forms $\lambda_i,$ together with a splitting 
	$$TM=\mathcal R\oplus\xi$$ 
	of the tangent bundle, satisfying the following conditions:
	\begin{enumerate}
	    \item $\xi:=\bigcap_{i}\ker\lambda_i;$
        \item for every $i,$ the restriction $d\lambda_i|_\xi$ is non-degenerate;
        \item for every $i,$ one has $\ker d\lambda_i=\mathcal R.$
	\end{enumerate}
\end{definition} 
	\begin{remark}
	The linear independence of the $\lambda_i$ implies that $\xi$ has constant rank $2n.$ Therefore, condition $(ii)$ is equivalent to $(d\lambda_i)^n|_\xi\neq 0,$ or, in other words, to $(\xi,d\lambda_i)$ being a symplectic bundle over $M.$ 
\end{remark}
A manifold endowed with such a structure is called a $q$-\emph{contact manifold} and denoted by $(M,\vec{\lambda},\mathcal R\oplus\xi),$ or simply by $M$ when the context permits. We call the collection $\{\lambda_i \}$ an \emph{adapted coframe} for the $q$-contact structure, and the $q$-form
$$\lambda:=\lambda_1\wedge\cdots\wedge\lambda_q\neq0$$
is called the \emph{characteristic form}. The bundles $\mathcal R$ and $\xi$ are called the \emph{Reeb distribution} and $q$-\emph{contact distribution}, respectively. The elements of $\xi$ will be called \emph{horizontal vector fields}.
\begin{definition}(Uniform $q$-contact structures)\label{D3} An adapted coframe $\vec{\lambda}=\{\lambda_i\}$ (and the $q$-contact structure it defines) is called uniform if it satisfies
	$$d\lambda_i=d\lambda_1$$
	for all $1\leq i\leq q.$
	A $q$-contact foliation $\mathcal F$ is said to be uniform if its adapted coframe is a reparameterisation of a uniform coframe. A manifold endowed with such structure is called a \emph{uniform} $q$-\emph{contact manifold}.
\end{definition}
\begin{proposition}\cite{Almeida}\label{P1} There is a unique collection of linearly independent vector fields $R_1,...,R_q$ tangent to $\mathcal R$ satisfying, for all $i,j=1,...,q,$ the relations
	\begin{enumerate}
	    \item  $\lambda_i(R_j)=\delta_{i}^j;$
        \item  $[R_i,R_j]=0;$
        \item  $\mathcal R=\operatorname{span}\{R_1,...,R_q \}$.
	\end{enumerate}
\end{proposition}
\begin{remark}
	In fact, in Proposition \ref{P1}, if the vector fields $R_1,...,R_q$ satisfy conditions (i) and (iii), then condition (ii) is automatically satisfied as well. 
\end{remark}
\begin{proof}
	On the one hand, we can see that for any $R_i,R_j$ and $\lambda_k$
	\begin{align*}
		0=\mathcal L_{R_i}(i_{R_j}\lambda_k)=i_{[R_i,R_j]}\lambda_k+i_{R_j}\mathcal L_{R_i}\lambda_k=i_{[R_i,R_j]}\lambda_k,
	\end{align*}
	thus, $[R_i,R_j]\in \xi.$ On the other hand, since
	$$0=\mathcal L_{R_i}(i_{R_j}d\lambda_k)=i_{[R_i,R_j]}d\lambda_k+i_{R_j}\mathcal L_{R_i}(d\lambda_k)=i_{[R_i,R_j]}d\lambda_k,$$
	we know that $[R_i,R_j]=0.$
\end{proof}
\begin{remark}
	By Proposition \ref{P1}, we know that for a uniform $q$-contact manifold, its Reeb distribution is Frobenius integrable.
\end{remark}
Hence, if $X$ is a vector field on a uniform $q$-contact manifold $M,$ we can write
$$X=(i_X\lambda_1)R_1+\cdots(i_X\lambda_1)R_q+\tilde X,$$
for a unique horizontal vector field $\tilde X$ (to verify this, apply $\lambda_i$ to both sides of the equation). We can decompose the cotangent bundle $T^*M$ by taking the annihilators of both components
$$T^*M=\mathcal R^o\oplus\xi^o.$$
The elements of $\mathcal R^o$ (hence the $1$-forms $\beta$ such that $\beta(R_i)=0,i=1,...,q$) are called  \emph{semi-basic forms}. Thus every differential $1$-form $\beta$ on $M$ can be written as 
$$\beta=(i_{R_1}\beta)\lambda_1+\cdots+(i_{R_q}\beta)\lambda_q+\hat\beta,$$
where $\hat\beta$ is a unique semi-basic form on $M$ (to verify this, evaluate both sides in $R_i$).
\begin{definition}\label{D5}(Reeb vector fields) The (unique) linearly independent vector fields $R_i,i=1,...,q$ from Proposition \ref{P1} are called the \emph{Reeb vector fields} of the $q$-contact structure.
\end{definition}
\begin{example}
	On \(\mathbb{R}^{2n+q}\) with coordinates \((x_1, y_1, \dots, x_n, y_n, z_1, \dots, z_q)\), there exists a simple \(q\)-contact structure given by the following \(q\) $1$-forms:
	\[\lambda_i := dz_i + \sum_{j=1}^n x_j dy_j,\quad i=1,\dots,q.\]
	Their Reeb vector fields are \(R_i = \partial_{z_i}\), and the Reeb distribution is \(\mathcal{R} = \operatorname{span}\{\partial_{z_1}, \dots, \partial_{z_q}\}\). The \(q\)-contact distribution is
	\[\xi = \operatorname{span}\{\partial_{x_1}, Y_1, \dots, \partial_{x_n}, Y_n\},\]
	where \(Y_i := \partial_{y_i} - x_i \sum_j \partial_{z_j}\) for each \(i\). Moreover, the volume form \((d\lambda_i)^n \wedge \lambda_1 \wedge \cdots \wedge \lambda_q\) coincides with the standard volume form on \(\mathbb{R}^{2n+q}\), namely
	\[\Omega = dx_1 \wedge dy_1 \wedge \cdots \wedge dx_n \wedge dy_n \wedge dz_1 \wedge \cdots \wedge dz_q.\]
\end{example}
	As is well known, on a contact manifold $(M,\eta)$, the Hamiltonian vector field $X_H$ is given by the formula:
		\begin{align*}
		\begin{cases}
			\text{}\eta(X_H)=-H,\\
			\text{}	i_{X_H}d\eta=dH-\mathcal R(H)\eta.
		\end{cases}
	\end{align*}
    \begin{example}
        Let $M = \mathbb{R}^3$ with coordinates $(q, p, z)$, and let
$\eta = dz - p\,dq$. Then $\eta$ is a contact structure on $M$, so $(M,\eta)$ is a contact manifold. Let
$H = \frac{p^2+q^2}{2}$, then its corresponding Hamiltonian system is given by:
\begin{align*}
\begin{cases} 
\dot{q} = p, \\
\dot{p} = -q,\\
\dot{z} = \frac{p^2-q^2}{2}. 
\end{cases}
\end{align*}
    \end{example}
	
Similarly, on a uniform $q$-contact manifold $(M,\vec{\lambda},\mathcal R\oplus\xi)$, for any smooth function $H$, there also exists a unique Hamiltonian vector field $X_H$ corresponding to that function, which is defined as follows.  
	\begin{theorem}\label{T2}
		Let $(M,\vec{\lambda},\mathcal R\oplus\xi)$ be a uniform $q$-contact manifold satisfying Proposition \ref{P1}.  Then, there is a unique vector field
		$X_H$ defined by
		\begin{align}\label{CH}
			\begin{cases}
				\text{}\lambda_i(X_H)=-H,\\
				\text{}	i_{X_H}d\lambda_1=dH-dH(R_1)\lambda_1-\cdots-dH(R_q)\lambda_q.
			\end{cases}
		\end{align}
	\end{theorem}
	\begin{proof}
	First of all, let \( H \) be a smooth function on \( M \). Then, we compute:
	\[
	\left(dH - dH(R_1)\lambda_1 - \cdots - dH(R_q)\lambda_q\right)(R_i) = dH(R_i) - dH(R_i) = 0, \quad i = 1, \dots, q.
	\]
	This shows that the $1$-form \( dH - dH(R_1)\lambda_1 - \cdots - dH(R_q)\lambda_q \) is semi-basic. Hence, there exists a unique horizontal vector field
	\[
	\widetilde{X_H} := (d\lambda_1)^\#\left(dH - dH(R_1)\lambda_1 - \cdots - dH(R_q)\lambda_q\right)
	\]
	such that
	\[
	d\lambda_1(\widetilde{X_H}, \cdot) = dH - dH(R_1)\lambda_1 - \cdots - dH(R_q)\lambda_q.
	\]
	Since \( d\lambda_1(R_i, \cdot) = 0 \) for all \( i \), we also have
	\[
	d\lambda_1(f_1 R_1 + \cdots + f_q R_q + \widetilde{X_H}, \cdot) = dH - dH(R_1)\lambda_1 - \cdots - dH(R_q)\lambda_q
	\]
	for any smooth functions \( f_1, \dots, f_q \) on \( M \).
	
	Furthermore, since \( \widetilde{X_H} \) is horizontal, we find that
	\[
	\lambda_i(f_1 R_1 + \cdots + f_q R_q + \widetilde{X_H}) = f_i \lambda_i(R_i) + \lambda_i(\widetilde{X_H}) = f_i,
	\]
	where we used the property \( \lambda_i(R_j) = \delta_{i}^j \) and \( \lambda_i(\widetilde{X_H}) = 0 \) because \( \widetilde{X_H} \) is horizontal.
	
	Taking \( f_i = -H \), we construct the vector field
	\[
	X_H := -H R_1 - \cdots - H R_q + \widetilde{X_H},
	\]
	which satisfies both of the desired conditions. By the construction above, this vector field is unique.
	\end{proof}
\begin{definition}\label{DeH}
In Theorem \ref{T2}, for a smooth function \( H \) on a uniform \( q \)-contact manifold, the associated vector field \( X_H \) is called the \( q \)-contact Hamiltonian vector field (or simply the  Hamiltonian vector field). The corresponding differential equation
\[
\dot{z} = X_H(z)
\]
is referred to as the \( q \)-contact Hamiltonian equation.
\end{definition}
\begin{example}
	Let $M = \mathbb{R}^4$ with coordinates $(z_1, z_2, q, p)$, and let
	$\lambda_1 = dz_1 - p\,dq$, $\lambda_2 = dz_2 + q\,dp$. Clearly $d\lambda_1=d\lambda_2,\mathcal R=\operatorname{span}\{z_1,z_2\},\xi=\operatorname{span}\{q,p\}$, hence $(M,\vec{\lambda}=(\lambda_1,\lambda_2),\mathcal R\oplus\xi)$ form a uniform 2-contact manifold. Let
	$H = \frac{p^2+q^2}{2}$, then its corresponding Hamiltonian system is given by:
	\begin{align*}
		\begin{cases} 
			\dot{q} = p, \\
			\dot{p} = -q,\\
			\dot{z}_1 = \frac{p^2-q^2}{2}, \\
			\dot{z}_2 = -\frac{p^2-q^2}{2}. 
		\end{cases}
	\end{align*}
\end{example}
\begin{remark}\label{PL}
Moreover, we can compute, for any \( 1 \leq i \leq q \),
\begin{align*}
	\mathcal L_{X_H} \lambda_i 
	= d(i_{X_H} \lambda_i) + i_{X_H} d\lambda_i 
	&= -dH + dH - dH(R_1)\lambda_1 - \cdots - dH(R_q)\lambda_q \\
	&= -dH(R_1)\lambda_1 - \cdots - dH(R_q)\lambda_q.
\end{align*}
Therefore, the equation \eqref{CH} is equivalent to
\begin{align} \label{CH2}
	\begin{cases}
		\lambda_i(X_H) = -H, \\
		\mathcal L_{X_H} \lambda_1 = -dH(R_1)\lambda_1 - \cdots - dH(R_q)\lambda_q.
	\end{cases}
\end{align}
\end{remark}

Now, we define a $q$-contact bracket on a uniform $q$-contact manifold.
\begin{definition}\label{D2}
	Let $(M,\vec{\lambda},\mathcal R\oplus\xi)$ be a uniform $q$-contact manifold satisfying Proposition \ref{P1}. We define the $q$-contact bracket $\{\cdot,\cdot \}:C^\infty(M)\times C^\infty(M)\rightarrow C^\infty(M)$ as:
	\begin{align}\label{dissipation}
		(f,g)\rightarrow \{f,g\}=\lambda_1([X_f,X_g])=\mathcal L_{X_f}(\lambda_1(X_g))-(\mathcal L_{X_f}\lambda_1)(X_g)=-X_f(g)-g\sum_{i=1}^qR_i(f),
	\end{align}
	where $[X_f,X_g]$ is the standard Jacobi--Lie bracket of the vector fields $X_f$ and $X_g.$
\end{definition}
\begin{definition}
\begin{enumerate}
    \item As a particular case of \eqref{dissipation}, we have 
	$$X_H(H)=-H\sum_{i=1}^q R_i(H),$$
	which we call the \emph{dissipation law} of the Hamiltonian $H$. In this sense, we will say that a $f$ is a \emph{dissipated quantity} of the Hamiltonian vector field $X_H$ if $\{H,f\}=0$, that is, $X_H(f)=-f\sum_{i=1}^qR_i(H)$.
    \item Similarly, we say that $g\in C^\infty(M)$ is a \emph{conserved quantity} if $X_H(g)=0.$
\end{enumerate}
\end{definition}
\begin{remark}
	We can verify that if $f_1$ and $f_2$ are dissipated quantities of $X_H$ and $f_2$ is nonvanishing, then $f_1/f_2$ is a conserved quantity of $X_H$.
\end{remark}
\begin{proof}
	In fact,
	$$X_H(f_1/f_2)=\frac{f_2X_H(f_1)-f_1X_H(f_2)}{f_2^2}=\frac{-f_2f_1\sum_{i=1}^qR_i(H)+f_1f_2\sum_{i=1}^qR_i(H)}{f_2^2}=0.$$
\end{proof}
\begin{remark}\label{R3}
	We can verify that $$\lambda_i([X_f,X_g])=\lambda_j([X_f,X_g]),i,j=1,...,q,$$ which indicates that the contact bracket we defined is independent of the choice of $\lambda_i$.
\end{remark}
\begin{proof}
In fact, we know that
\[
\lambda_i(X_f) = \lambda_j(X_f) = -f, \quad \text{for all } 1 \leq i, j \leq q.
\]
Therefore, we compute:
\begin{align*}
	0 =\mathcal L_{X_g}((\lambda_i - \lambda_j)(X_f))
	&= \left(\mathcal L_{X_g}(\lambda_i - \lambda_j)\right)(X_f) + (\lambda_i - \lambda_j)( \mathcal L_{X_g} X_f) \\
	&= \left(d(i_{X_g}(\lambda_i - \lambda_j))\right)(X_f) + \left(i_{X_g}(d(\lambda_i - \lambda_j))\right)(X_f) + (\lambda_i - \lambda_j)(\mathcal L_{X_g} X_f) \\
	&= \left(d(g - g)\right)(X_f) + \left(X_g \lrcorner (d\lambda_1 - d\lambda_1)\right)(X_f) + (\lambda_i - \lambda_j)( \mathcal L_{X_g} X_f) \\
	&= (\lambda_i - \lambda_j)( \mathcal L_{X_g} X_f),
\end{align*}
which implies that
\[
\lambda_i([X_f, X_g]) = \lambda_j([X_f, X_g]), \quad \text{for all } i, j = 1, \dots, q.
\]
.
\end{proof}
Now, we present an additional property of the $q$-contact bracket.
	
	\begin{proposition} \label{P6}
	Let \( (M, \vec{\lambda}, \mathcal{R} \oplus \xi) \) be a uniform \( q \)-contact manifold satisfying Proposition \ref{P1}. If \( X \) is a vector field such that \( \lambda_i(X) = -f \) for all \( i = 1, \dots, q \), then the following identity holds:
	\begin{align}
		\{H, f\} = -\lambda_i([X_H, X]) = ( \mathcal L_X \lambda_i)(X_H) + X(H).
	\end{align}
	\end{proposition}
	
	\begin{proof}
	If \( \lambda_i(X) = -f \) for all \( i = 1, \dots, q \), then
	\[
	\lambda_i(X - X_f) = 0, \quad \text{for all } i = 1, \dots, q,
	\]
	which implies that \( X - X_f \in \bigcap_{i=1}^q \ker \lambda_i \).
	
	By Remark \ref{PL}, we know that for each \( i = 1, \dots, q \),
	\begin{align*}
		\mathcal L_{X_H} \lambda_i 
		= d(i_{X_H} \lambda_i) + i_{X_H} d\lambda_i 
		= -dH(R_1)\lambda_1 - \cdots - dH(R_q)\lambda_q,
	\end{align*}
	so we obtain
	\[
	(\mathcal L_{X_H} \lambda_i)(X_f) = (\mathcal L_{X_H} \lambda_i)(X).
	\]
	
	Therefore, using Remark \ref{R3}, we compute for any \( i = 1, \dots, q \):
	\begin{align*}
		\{H, f\} 
		&= \lambda_i([X_H, X_f]) \\
		&=\mathcal L_{X_H}(\lambda_i(X_f)) - ( \mathcal L_{X_H} \lambda_i)(X_f) \\
		&= \mathcal L_{X_H}(\lambda_i(X)) - (\mathcal L_{X_H} \lambda_i)(X) \\
		&= \lambda_i([X_H, X]).
	\end{align*}
	
	Applying Cartan's magic formula again, for each \( i = 1, \dots, q \), we have:
	\begin{align*}
		\lambda_i([X_H, X]) 
		= -X(\lambda_i(X_H)) + (\mathcal L_X \lambda_i)(X_H) 
		= X(H) + (\mathcal L_X \lambda_i)(X_H).
	\end{align*}
\end{proof}
\begin{remark}\label{R7}
	We can see that if a vector field $X$ satisfies the conditions in Proposition \ref{P6}, then 
	$$\lambda_1([X_H,X])=\cdots=\lambda_q([X_H,X]).$$
\end{remark}
\begin{definition}	Let \( (M, \vec{\lambda}, \mathcal{R} \oplus \xi) \) be a uniform \( q \)-contact manifold satisfying Proposition \ref{P1}. For a Hamiltonian vector field $X_H,$
	 vector field $Y\in\mathfrak{X}(M)$ is said to be a \emph{dynamical symmetry} if $\mathcal L_YX_H=0.$
\end{definition}
\begin{proposition}\label{P8}
	If $Y$ is a dynamical symmetry of $X_H$ and $\lambda_1(Y)=\cdots=\lambda_q(Y)$, then 
	\begin{enumerate}
	    \item $\lambda_1(Y)$ is a dissipated quantity of $X_H$.
        \item \( H \) is a conserved quantity of \( Y \) if and only if  \( X_H \) annihilates all $1$-forms \( \mathcal L_Y \lambda_1, \dots,\mathcal L_Y \lambda_q \).
	\end{enumerate}
\end{proposition}
\begin{proof}
	Let $\lambda_1(Y)=-f,$ then by Proposition \ref{P6} and Remark \ref{R7}, we know that 
	$$\{H,f\}=-\{H,\lambda_1(Y)\}=-\lambda_1([X_H,Y])=\cdots=-\lambda_q([X_H,Y]).$$
Hence, if $Y$ is a dynamical symmetry, i.e., $[X_H,Y]=0$, $\{H,\lambda_1(Y)\}=0$, then $\lambda_1(Y)$ is a dissipated quantity of $X_H.$ Moreover, since for any $i\in\{1,...,q\}$,
\begin{align*}
	0=-\lambda_i([X_H,Y])&=\lambda_i([Y,X_H])=\mathcal L_Y(\lambda_i(X_H))-( \mathcal L_Y\lambda_i)(X_H)=-Y(H)-( \mathcal L_Y\lambda_i)(X_H),
\end{align*}
we know that $Y(H)=0$ if and only if $(\mathcal L_Y\lambda_i)(X_H)=0,\forall i=1,...,q.$
\end{proof}
\begin{remark}\label{R8}
	If $Y$ satisfies the conditions in Proposition \ref{P8} and $i_Yd\lambda_1=0,$ then  \( H \) is a first integral of \( Y \) if and only if $\lambda_1(Y)$ is a conserved quantity of \( X_H \).
\end{remark}
\begin{proof}
	By the proof of Proposition \ref{P8}, we know that
	$$Y(H)=-(\mathcal L_Y\lambda_i)(X_H)=-(i_Yd\lambda_i+d(\lambda_i(Y)))(X_H)=-X_H(\lambda_i(Y)),$$
	which implies the result.
\end{proof}
\section{q-Contact Lagrangian System}
Let $Q$ be an $n$-dimensional manifold, and we consider its
tangent bundle $TQ$ and the extended phase space $TQ\times \mathbb R^q$. We denote by $(q^i,\dot q^j,z^1,...,z^q)$
the bundle coordinates on  $TQ\times \mathbb R^q$ and consider the natural extension to
 $TQ\times \mathbb R^q$ of the canonical vertical endomorphism $S$ on $TQ$ locally deﬁned
by the $(1, 1)$-tensor ﬁeld
\begin{align*}
	S=\frac{\partial}{\partial \dot q^i}\otimes dq^i.
\end{align*}
Let  $L:TQ\times \mathbb R^q\rightarrow \mathbb R$ be a Lagrangian function, where $Q$ is the $n$-dimensional configuration manifold of a mechanical system. Then, $L=L(q^i,\dot q^i,z^k)$, where $(q^i)$ are coordinates in $Q$, $(q^i,\dot q^i)$ are the induced bundle coordinates in $TQ$ and $z^k$ are global coordinates in $\mathbb R^q.$

We say that $L$ is \emph{regular} if the Hessian matrix with respect to the velocities
\begin{align}
	W_{ij}=\left(\frac{\partial^2L}{\partial \dot q^i\partial \dot q^j}\right)
\end{align}
is regular.

From $L$, and using the canonical endomorphism $S$ on $TQ,$ we can construct a $1$-form $\eta_L$  defined by
\begin{align}\label{SDL}
	\eta_L=S^*(dL)=\frac{\partial L}{\partial\dot q^i}dq^i,
\end{align}
where now, by an abuse of notation, $S$ and $S^*$ are the natural extensions of $S$ and its adjoint operator $S^*$ to $TQ\times \mathbb R^q.$ Therefore, we have 
\begin{align}
	\eta_L&=\frac{\partial L}{\partial\dot q^i}dq^i,\\
	d\eta_L&=\frac{\partial^2 L}{\partial\dot q^i\partial q^j}dq^j\wedge dq^i+\frac{\partial^2 L}{\partial\dot q^i\partial\dot q^k}d\dot q^k\wedge dq^i+\frac{\partial^2 L}{\partial\dot q^i\partial z^m}dz^m\wedge dq^i.
\end{align}
Consider the $1$-forms
\begin{align}
 \lambda^L_i=dz^i-\eta_L=dz^i-\frac{\partial L}{\partial\dot q^j}dq^j,\quad i=1,...,q.
\end{align}
We know that $d\lambda^L_1=\cdots=d\lambda^L_q$ and if $L$ is regular, then
\begin{align}
	\lambda^L_1\wedge\cdots\wedge\lambda^L_q\wedge (\Omega_L)^n=	\lambda^L_1\wedge\cdots\wedge\lambda^L_q\wedge(d\lambda^L_i)^n\neq0,\quad i=1,...,q.
\end{align}
Here and below, we will always assume that $L$ is regular.

 Define the following vector fields,
\begin{align}
	R_k=\frac{\partial}{\partial z^k}-W^{ij}\frac{\partial^2L}{\partial \dot q^i\partial z^k}\frac{\partial}{\partial\dot q^j},
\end{align}
where $(W^{ij})$ is the inverse of the Hessian matrix of $L$ with respect to the velocities. We can also verify that 
\begin{align}
	\lambda^L_i(R_k)=\delta_k^i,\quad i_{R_j}d\lambda_1^L=0,\quad i,j,k=1,...,q.
\end{align}
Let $\mathcal R=\operatorname{span}\{R_1,...,R_q\},$ $\xi=\bigcap_{i=1}^q\ker\lambda_i^L,$ then we know that $(TQ\times \mathbb R^q,\vec\lambda=(\lambda_1^L,...,\lambda_q^L),\mathcal R\oplus \xi)$ is a unique $q$-contact manifold.

Also we deﬁne the \emph{Liouville vector ﬁeld} $\Delta$ on $TQ \times \mathbb R^q$ by the local expression
\begin{align}
	\Delta=\dot q^i\frac{\partial}{\partial \dot q^i},
\end{align}
and the \emph{energy function} of the system is defined as
\begin{align}
	E_L=\Delta(L)-L=\dot q^i\frac{\partial L}{\partial \dot q^i}-L.
\end{align}
We say that $(TQ\times\mathbb R^q,\vec{\lambda},E_L)$ is a $q$-\emph{contact Lagrangian system}.
 
Moreover, in local coordinates, the Hamiltonian vector field $X_{E_L}$ of the energy function $E_L$ on the unique $q$-contact manifold $(TQ\times \mathbb R^q,\vec\lambda=(\lambda_1^L,...,\lambda_q^L),\mathcal R\oplus \xi)$ can be described as follows:  
\begin{align}\label{i0}
	i_{X_{E_L}}d\lambda_1^L=dE_L-R_1(E_L)\lambda_1^L-\cdots-R_q(E_L)\lambda_q^L,
\end{align}
thus, we can get that
\begin{align}\label{i6}
X_{E_L} = \dot q^i \frac{\partial }{\partial q^i} + W^{ik} \left( \frac{\partial L}{\partial q^k} 
- \frac{\partial^2 L}{\partial q^j \partial \dot q^k}\dot q^j +\sum_{l=1}^q\left( - L \frac{\partial^2 L}{\partial z^l \partial \dot q^k} 
+ \frac{\partial L}{\partial z^l} \frac{\partial L}{\partial \dot q^k} \right)\right) \frac{\partial}{\partial \dot q^i} + L \sum_{l=1}^q\frac{\partial}{\partial z^l}.
\end{align}
In fact, we can calculate
\begin{align}
	i_{X_{E_L}}d\lambda_1^L&=-i_{ \dot q^i \frac{\partial }{\partial q^i}}\left(\frac{\partial^2 L}{\partial\dot q^i\partial q^j}dq^j\wedge dq^i+\frac{\partial^2 L}{\partial\dot q^i\partial\dot q^k}d\dot q^k\wedge dq^i+\frac{\partial^2 L}{\partial\dot q^i\partial z^m}dz^m\wedge dq^i\right)\nonumber\\
	&\quad-i_{ W^{ik} \left( \frac{\partial L}{\partial q^k} 
		- \frac{\partial^2 L}{\partial q^j \partial \dot q^k}\dot q^j +\sum_{l=1}^q\left( - L \frac{\partial^2 L}{\partial z^l \partial \dot q^k} 
		+ \frac{\partial L}{\partial z^l} \frac{\partial L}{\partial \dot q^k} \right)\right) \frac{\partial}{\partial \dot q^i}}\left(\frac{\partial^2 L}{\partial\dot q^i\partial\dot q^k}d\dot q^k\wedge dq^i\right)\nonumber\\
		&\quad -i_{L \sum_{l=1}^q\frac{\partial}{\partial z^l}}\left(\frac{\partial^2 L}{\partial\dot q^i\partial z^m}dz^m\wedge dq^i\right).\label{i5}
\end{align}
Notice that
\begin{align}
	&i_{ \dot q^i \frac{\partial }{\partial q^i}}\left(\frac{\partial^2 L}{\partial\dot q^i\partial q^j}dq^j\wedge dq^i+\frac{\partial^2 L}{\partial\dot q^i\partial\dot q^k}d\dot q^k\wedge dq^i+\frac{\partial^2 L}{\partial\dot q^i\partial z^m}dz^m\wedge dq^i\right)=\dot q^i\left(\frac{\partial^2 L}{\partial\dot q^i\partial q^j}-\frac{\partial^2 L}{\partial q^i\partial\dot q^j}\right)dq^j\nonumber\\
	&\quad\quad\quad\quad\quad\quad\quad\quad\quad\quad\quad\quad\quad\quad\quad\quad\quad\quad\quad\quad+\left(\dot q^i\frac{\partial^2 L}{\partial \dot q^i\partial\dot q^j}\right)d\dot q^j+\left(\dot q^i\frac{\partial^2 L}{\partial \dot q^i\partial z^m}\right)dz^m,\label{i1}\\
		&i_{ W^{ik} \left( \frac{\partial L}{\partial q^k} 
		- \frac{\partial^2 L}{\partial q^j \partial \dot q^k}\dot q^j +\sum_{l=1}^q\left( - L \frac{\partial^2 L}{\partial z^l \partial \dot q^k} 
		+ \frac{\partial L}{\partial z^l} \frac{\partial L}{\partial \dot q^k} \right)\right) \frac{\partial}{\partial \dot q^i}}\left(\frac{\partial^2 L}{\partial\dot q^i\partial\dot q^k}d\dot q^k\wedge dq^i\right)\nonumber\\
		&\quad\quad\quad\quad\quad\quad\quad\quad\quad\quad\quad\quad\quad=-\left( \frac{\partial L}{\partial q^i}-\frac{\partial^2 L}{\partial q^j\partial\dot q^i}\dot q^j+\sum_{l=1}^q\left(-L\frac{\partial L^2}{\partial z^l\partial\dot q^i+\frac{\partial L}{\partial z^l}\frac{\partial L}{\partial\dot q^i}}\right)\right)dq^i,\label{i2}\\
		&\quad\quad\quad\quad\quad\quad i_{L \sum_{l=1}^q\frac{\partial}{\partial z^l}}\left(\frac{\partial^2 L}{\partial\dot q^i\partial z^m}dz^m\wedge dq^i\right)=-L\left(\sum_{l=1}^q\frac{\partial L^2}{\partial \dot q^i\partial z^l}\right)dq^i,\label{i3}\\
		&dE_L-\sum_{l=1}^qR_l(X_{E_L})\lambda^L_l=\dot q^i\frac{\partial^2L}{\partial q^j\partial\dot q^i}dq^j+\dot q^i\frac{\partial^2L}{\partial \dot q^j\partial\dot q^i}dq^j+\dot q^i\frac{\partial^2L}{\partial z^l\partial\dot q^i}dz^l-\frac{\partial L}{\partial q^i}dq^i-\frac{\partial L}{\partial z^l}dz^l\nonumber\\
		&\quad\quad\quad\quad\quad\quad\quad\quad\quad+\sum_{l=1}^q\left(\frac{\partial L}{\partial z^l}\left(dz^l-\frac{\partial L}{\partial\dot q^i}dq^i\right)\right).\label{i4}
\end{align}
Hence, we know that $\eqref{i1},\eqref{i2},\eqref{i3}$ and $\eqref{i4}$ implies \eqref{i0}, thus we can get that $\eqref{i6}.$
Moreover, we can get 
\begin{align*}
	S(X_{E_L}) = \Delta,\quad R_i(E_L)=-\frac{\partial L}{\partial z^i},\quad \lambda^L_i(X_{E_L})=L-\dot q^j\frac{\partial L}{\partial \dot q^j}=-E_L,\quad dz^i(X_{E_L})=L.
\end{align*}
\section{$q$-Contact Noether Theorem and Symmetries}
As we describe in the introduction, the geometric description of Noether's Theorem for a Lagrangian $L:TQ\rightarrow \mathbb R$ states that the invariance condition $X^c(L)=0$ for the 
lift $X^c$ to $TQ$ of a vector field $X$ is equivalent to $X^v(L)$ being a conserved quantity, where $X^v$ is a vertical field on $TQ$ over $Q,$ canonically associated with $X.$

Due to the difference between symplectic and $q$-contact Hamiltonian dynamics, the Noether Theorem undergoes a radical transformation in the transition from a Lagrangian system to a $q$-contact Lagrangian system: the invariance condition will be more complicated and it does not yield conserved quantities but rather dissipated quantities.

Consider the vector field $X$ on $TQ\times \mathbb R^q$. In local coordinates $(q^j,\dot q^i,z^1,...,z^q)$ on $TQ\times \mathbb R^q$, it takes the form
\begin{align}\label{XF}
	X=F_i\frac{\partial}{\partial q^i} +G_j\frac{\partial}{\partial \dot q^j}+H_k\frac{\partial}{\partial z^k},
\end{align}
where
$$F_i,G_j,H_k\in C^\infty(TQ\times \mathbb R^q).$$
Using the notation 
$$X^v=S(X)=F_i\frac{\partial}{\partial\dot q_i},$$ 
if we let $f=\lambda_i^L(X),$ then
$$f=(dz^i-\eta_L)(X)=(H_i-S^*(dL)(X))=(H_i-X^vL)=\left(H_i-\sum_{j=1}^qF_j\frac{\partial L}{\partial\dot q_j}\right).$$
Now we can see that 
\begin{align}
	d(S^*(dL))(X_{E_L},X)&=i_Xi_{X_{E_L}}(d(S^*dL))\nonumber\\
	&=i_X(\mathcal L_{X_{E_L}}(S^*dL)-di_{X_{E_L}}(S^*dL))\nonumber\\
	&=i_X\mathcal L_{X_{E_L}}(S^*dL)-X(S^*dL(X_{E_L}))\nonumber\\
	&=\mathcal L_{X_{E_L}}(i_X(S^*dL))-i_{[X_{E_L},X]}S^*dL-X(S^*dL(X_{E_L}))\nonumber\\
	&=X_{E_L}(X^vL)-X(\Delta L)-[X_{E_L},X]^vL.\label{shi1}
\end{align}
Moreover, utilizing the relations
$$	i_{X_{E_L}}d\lambda_1^L=dE_L-R_1(E_L)\lambda_1^L-\cdots-R_q(E_L)\lambda_q^L$$
and $d\lambda_i^L=-d\eta_L=-d(S^*dL)$, we obtain 
\begin{align}
	d(S^*(dL))(X_{E_L},X)&=i_Xi_{X_{E_L}}(d(S^*dL))\nonumber\\
	&=-X(E_L)+\sum_{i=1}^qR_i(E_L)\lambda_i^L(X)\nonumber\\
	&=-X(\Delta L)+X(L)+\sum_{i=1}^qR_i(E_L)\lambda_i^L(X)\nonumber\\
	&=-X(\Delta L)+X(L)+\sum_{i=1}^qR_i(E_L)(H_i-X^vL),\label{shi2}
\end{align}
thus, by \eqref{shi1} and \eqref{shi2}, we get that 
$$X_{E_L}(X^vL)-[X_{E_L},X]^vL=X(L)+\sum_{i=1}^qR_i(E_L)(H_i-X^vL),$$
which means that
\begin{align}\label{EL}
	\{E_L,X^vL\}=-X(L)-[X_{E_L},X]^vL-\sum_{i=1}^qR_i(E_L)H_i=-X(L)-[X_{E_L},X]^vL+\sum_{i=1}^q\frac{\partial L}{\partial z^i}H_i.
\end{align} 
Summarizing, we have
\begin{theorem}\label{GN}(Noether-Type Theorem)
	Given a vector field $X\in\mathfrak{X}(TQ\times\mathbb R^q)$ in the form \eqref{XF},  if $$-X(L)-[X_{E_L},X]^vL+\sum_{i=1}^q\frac{\partial L}{\partial z^i}H_i=0,$$  then the function $X^v(L)$ is a dissipated quantity for the $q$-contact Lagrangian system defined by $L.$
\end{theorem}

Consider the expression \eqref{EL} for the particular case where the vector field $X$ on $TQ\times \mathbb R^q$ is the complete lift $Y^c$ of a field $Y\in \mathfrak (Q).$ If $Y$ is a vector field on $Q$ given locally by
	$$Y=Y^i\frac{\partial }{\partial q^i},\quad Y^i\in C^\infty(Q),$$
	its complete lift is given by 
	$$Y^c=Y^i\frac{\partial}{\partial q^i}+\dot q^j\frac{\partial Y^i}{\partial q^j}\frac{\partial}{\partial \dot q^i},$$
	and its vertical lift is given by 
	$$Y^v=SY^c=Y^i\frac{\partial}{\partial\dot q^i}.$$
	These lifts can be defined geometrically \cite{de4,Yano}. Moreover, by \eqref{i6} we know that
	$$X_{E_L} = \dot q^i \frac{\partial }{\partial q^i} + W^{ik} \left( \frac{\partial L}{\partial q^k} 
	- \frac{\partial^2 L}{\partial q^j \partial \dot q^k}\dot q^j +\sum_{l=1}^q\left( - L \frac{\partial^2 L}{\partial z^l \partial \dot q^k} 
	+ \frac{\partial L}{\partial z^l} \frac{\partial L}{\partial \dot q^k} \right)\right) \frac{\partial}{\partial \dot q^i} + L \sum_{l=1}^q\frac{\partial}{\partial z^l}.$$
	Hence, by a straightforward computation, 
	$$[X_{E_L},Y^c]=A_i\frac{\partial}{\partial\dot q^i}+B_j\frac{\partial}{\partial z^j}\Longrightarrow [X_{E_L},Y^c]^v=S[X_{E_L},Y^c]=0.$$
	Now, \eqref{EL} becomes
	$$\{E_L,Y^v(L)\}=Y^c(L),$$
	which implies that $Y^v(L)$ is the dissipated quantity associated with the symmetry condition $Y^c(L)=0.$ We summarize this as
    
\begin{corollary}\label{C3}
	Given the complete lift $Y^c$ of a field $Y\in \mathfrak (Q)$, then
    $$\{E_L,Y^v(L)\}=Y^c(L),$$
    which implies that $Y^v(L)$ is the dissipated quantity associated with the symmetry condition $Y^c(L)=0$. This extends the Noether Theorem for conservative Lagrangian systems to the case of dissipative $q$-contact Lagrangian systems.
\end{corollary}
\begin{example}\label{E1}
	We consider a multi-parameter dependent $m$-contact Lagrangian function $$L=\frac{1}{2}v^2-\frac{1}{2}q^2-\gamma_1 z_1-\cdots-\gamma_m z_m,$$ on the manifold $T\mathbb R\times \mathbb R^m,$ where $\gamma_1,...,\gamma_m$ are constants.
	
	The $q$-contact structure is characterized by the following elements
	\begin{align*}
		\lambda_i^L&=dz_i-vdq,\quad d\lambda_i^L=dq\wedge dv,\quad i=1,...,m,\\
		R_j&=\frac{\partial}{\partial z_j},j=1,...,m,\quad S=\frac{\partial}{\partial v}\otimes dq,\quad \Delta=v\frac{\partial}{\partial v},\\
		E_L&=\Delta L-L=\frac{1}{2}v^2+\frac{1}{2}q^2+\gamma_1 z_1+\cdots+\gamma_m z_m.
	\end{align*}
	The Hamiltonian vector field $X_{E_L}$ arises from the dynamical equations
	\begin{align*}
		i_{X_{E_L}}\lambda_i^L&=-E_L,\quad i=1,...,m,\\
		i_{X_{E_L}}d\lambda_i^L&=i_{X_{E_L}}(dq\wedge dv)=dE_L-R_1(E_L)\lambda_1^L-\cdots-R_q(E_L)\lambda_m^L.
	\end{align*}
	Solving these equations yield
	$$X_{E_L}=-v\frac{\partial}{\partial q}+\left(q+v\sum_{i=1}^m\gamma_i\right)\frac{\partial}{\partial v}-L\sum_{i=1}^m\frac{\partial}{\partial z_i}.$$
	For any vector field $X$ which has the form 
	$$X=F\frac{\partial}{\partial q}+G\frac{\partial}{\partial v}+\sum_{i=1}^mH_i\frac{\partial}{\partial z_i},$$
	 by Theorem \ref{GN}, we know that 
	\begin{align*}	\{E_L,X^vL\}=\{E_L,vF\}&=-X(L)-[X_{E_L},X]^vL+\sum_{i=1}^m\frac{\partial L}{\partial z^i}H_i\\
		&=qF-(X_{E_L}(F)+G)v-\sum_{i=1}^m\gamma_iH_i,
	\end{align*}
Thus, in order to ensure that $\{E_L, X^v L\} = 0$, it suffices to find functions $F, G, H_1, \dots, H_m$ such that  
\begin{align}\label{equ}
qF - (X_{E_L}(F) + G)v - \sum_{i=1}^m \gamma_i H_i = 0.
\end{align}
For instance, by a straightforward calculation, we can verify that the functions  
\[
F = vK(q, v), \quad G = qK - \left(q + v\sum_{i=1}^m \gamma_i\right)K - vX_{E_L}(K), \quad \text{and} \quad H_i = 0 \quad \text{for } i = 1, \dots, m,
\]  
are solutions to equation \eqref{equ}, where $K(q, v)$ is an arbitrary function of $q$ and $v$.

However, we claim that there does not exist a vertical lifted vector field of the form $X = Y^v$, with $Y \in \mathfrak{X}(\mathbb{R})$, that satisfies the equation $\{E_L, Y^v(L)\} = 0$.  
Indeed, let $Y = W(q)\frac{\partial}{\partial q}$. Then its complete lift is given by  
\[
Y^c = W(q)\frac{\partial}{\partial q} + v\frac{\partial W}{\partial q}\frac{\partial}{\partial v}.
\]  
It is clear that  
\[
Y^c(L) = -W(q)q + v^2 \frac{\partial W}{\partial q} \neq 0.
\]  
Therefore, by Corollary~\ref{C3}, we conclude that  
\[
\{E_L, Y^v(L)\} = Y^c(L) \neq 0.
\]  
Hence, the claim follows.
\end{example}
\begin{definition}
	A vector field $Y\in \mathfrak{X}(TQ\times\mathbb R^q)$ on the $q$-contact Lagrangian system $(TQ\times\mathbb R^q,\vec{\lambda},E_L)$ is a Noether symmetry if $\mathcal L_YE_L=0$ and $\mathcal L_{Y}\lambda_i^L=0,i=1,...,q$.
\end{definition}
\begin{remark}
	Every Noether symmetry is also a dynamical symmetry, i.e., $[Y,X_{E_L}]=0$.
\end{remark}
\begin{proof}
	Indeed, to see $[Y,X_{E_L}]=0,$ it suffices to verify that
	$$i_{[Y,X_{E_L}]}d\lambda_1^L=0,\quad i_{[Y,X_{E_L}]}\lambda_j^L=0,j=1,...,q.$$
	Now
	$$i_{[Y,X_{E_L}]}\lambda_i^L=Y(\lambda_i^L(X_{E_L}))-(\mathcal L_Y\lambda_i^L)(X_{E_L})=Y(\lambda_i^L(X_{E_L}))=-Y(E_L)=0,\quad i=1,...,q,$$
	and
	\begin{align}
		i_{[Y,X_{E_L}]}d\lambda_1^L&=\mathcal L_Y(i_{X_{E_L}}d\lambda_1^L)-i_{X_{E_L}}\mathcal L_Yd\lambda_1^L\nonumber\\
		&=\mathcal L_Y(dE_L-R_1(E_L)\lambda_1^L-\cdots-R_q(E_L)\lambda_q^L)\nonumber\\
		&=-\sum_{i=1}^q[Y,R_i](E_L)\lambda_i^L.\label{LY}
	\end{align}
	On the other hand, we know that
	\begin{align*}
		0=\mathcal L_Y(\delta_{j}^k)=\mathcal L_Y(i_{R_j}\lambda_k^L)=i_{[Y,R_j]}\lambda_k^L+i_{R_j}\mathcal L_Y\lambda_k^L=i_{[Y,R_j]}\lambda_k^L, \quad j,k=1,...,q,
	\end{align*}
	which means that $[Y,R_j]\in\xi,j=1,...,q$. Moreover, since
	\begin{align*}
		i_{[Y,R_j]}d\lambda^L_1=\mathcal L_Y(i_{R_j}d\lambda^L_1)-i_{R_j}\mathcal L_Yd\lambda_1^L=0,\quad j=1,...,q,
	\end{align*}
	thus $[Y,R_j]=0,j=1,...,q$. Therefore, from \eqref{LY}, we obtain $	i_{[Y,X_{E_L}]}d\lambda_1^L=0,$ so $[Y,X_{E_L}]=0.$
\end{proof}
Now, we can state the following theorems.
\begin{theorem}
	Given a Hamiltonian $f\in C^\infty(TQ\times\mathbb R^q)$, if the Hamiltonian vector field $X_f$ is also a Noether symmetry, then $f$ is a dissipated quantity. Conversely, if $f$ is a dissipated quantity and $R_1(f)=\cdots=R_q(f)=0,$ then $X_f$ is a Noether symmetry.
\end{theorem}
\begin{proof}
	By Definition \ref{D2}, 
	$$\{f,E_L\}=-\mathcal L_{X_f}(E_L)-(\mathcal L_{X_f}\lambda^L_1)(X_{E_L})=0.$$
Hence, $f$ is a dissipated quantity. Conversely, if $f$ is a dissipated quantity and $R_1(f)=\cdots=R_q(f)=0$, then by Remark \ref{PL} and Definition \ref{D2},
\begin{align*}
	\mathcal L_{X_f}\lambda_i^L&=-\sum_{i=1}^qR_i(f)\lambda_i=0,\quad i=1,...,q\\
	0&=\{f,E_L\}=-\mathcal L_{X_f}(E_L)-(\mathcal L_{X_f}\lambda^L_1)(X_{E_L})=-\mathcal L_{X_f}E_L,
\end{align*}
which means that $X_f$ is a Noether symmetry.
\end{proof}
The following theorem establishes a connection between the dynamical symmetries  and the dissipative quantities.
\begin{theorem}\label{Th4}
	If the vector field $Y$ is a dynamical symmetry of $X_{E_L}$ and $\mathcal L_Y\left(\sum_{i=1}^qR_i(E_L)\right)=0$, then $\mathcal L_YE_L$ is a dissipative quantity. In addition, if $\lambda_1^L(Y)=\cdots=\lambda_q^L(Y),$ then $\mathcal L_Y(\lambda_i^L(Y)),i=1,...,q$ are all dissipative quantities.
\end{theorem}
\begin{proof}
	First of all, we know that
	$$0=i_{[X_{E_L},Y]}\lambda_i^L=\mathcal L_{X_{E_L}}(i_{Y}\lambda_i^L)-i_Y\mathcal L_{X_{E_L}}\lambda_i^L=X_{E_L}(\lambda_i^L(Y))+\sum_{j=1}^qR_j(H)\lambda_j^L(Y),$$
which leads to 
\begin{align}\label{XEL}
	X_{E_L}(\lambda_i^L(Y))=-\sum_{j=1}^qR_j(H)\lambda_j^L(Y).
\end{align}
Moreover, 
\begin{align*}
	\{E_L,\mathcal L_YE_L\}&=-X_{E_L}(\mathcal L_YE_L)-\mathcal L_YE_L\sum_{i=1}^qR_i(E_L)\\
	&=-\mathcal L_Y(X_{E_L}(E_L))+\mathcal L_YX_{E_L}(E_L)-\mathcal L_YE_L\sum_{i=1}^qR_i(E_L)\\
	&=\mathcal L_Y\left(E_L\sum_{i=1}^qR_i(E_L)\right)-\mathcal L_YE_L\sum_{i=1}^qR_i(E_L)\\
	&=\mathcal L_YE_L\sum_{i=1}^qR_i(E_L)+E_L\mathcal L_Y\left(\sum_{i=1}^qR_i(E_L)\right)-\mathcal L_YE_L\sum_{i=1}^qR_i(E_L)\\
	&=E_L\mathcal L_Y\left(\sum_{i=1}^qR_i(E_L)\right)=0.
\end{align*}
Furthermore, if $\lambda_1(Y)=\cdots=\lambda_q(Y),$ for any $i\in\{1,...,q\}$ by \eqref{XEL}, we have
\begin{align*}
	\{E_L,\mathcal L_Y(\lambda_i^L(Y))\}&=-X_{E_L}(\mathcal L_Y(\lambda_i^L(Y)))-\mathcal L_Y(\lambda_i^L(Y))\sum_{i=j}^qR_j(E_L)\\
	&=-\mathcal L_Y(X_{E_L}(\lambda_i^L(Y)))-\mathcal L_Y(\lambda_i^L(Y))\sum_{j=1}^qR_j(E_L)\\
	&=\mathcal L_Y\left(\sum_{k=1}^qR_k(E_H)\lambda_k(Y)\right)-\mathcal L_Y(\lambda_i^L(Y))\sum_{j=1}^qR_j(E_L)\\
	&=\lambda_i(Y)\mathcal L_Y\left(\sum_{k=1}^qR_k(E_L)\right)+\mathcal L_Y(\lambda_i^L(Y))\sum_{j=1}^qR_j(E_L)-\mathcal L_Y(\lambda_i^L(Y))\sum_{j=1}^qR_j(E_L)\\
	&=\lambda_i(Y)\mathcal L_Y\left(\sum_{k=1}^qR_k(E_H)\right)=0.
\end{align*}
\end{proof}
\begin{example}
Consider a 2-parameter dependent uniform 2-contact Lagrangian function 
$$L=\frac{v^2}{2}-z_1-z_2$$
on the manifold $T\mathbb R\times\mathbb R^2$.
The 2-contact structure is characterized by
\begin{align*}
		\lambda_i^L&=dz_i-vdq,\quad d\lambda_i^L=dq\wedge dv,\quad i=1,2,\\
		R_j&=\frac{\partial}{\partial z_j},j=1,2,\quad S=\frac{\partial}{\partial v}\otimes dq,\quad \Delta=v\frac{\partial}{\partial v},\\
		\xi&=\operatorname{span}\{\frac{\partial}{\partial q},\frac{\partial}{\partial v}\},\quad E_L=\Delta L-L=\frac{v^2}{2}+z_1+z_2.
	\end{align*}
    Similar to the calculation of Example \ref{E1}, we can get 
    $$X_{E_L}=v\frac{\partial}{\partial q}+\left(\frac{v^2}{2}-z_1-z_2\right)\frac{\partial}{\partial z_1}+\left(\frac{v^2}{2}-z_1-z_2\right)\frac{\partial}{\partial z_2},\quad  X_{E_L}(R_1(E_L)+R_2(E_L))=X_{E_L}(2)=0.$$
    Let vector field $Y:=X_{E_L}$, then $[Y,X_{E_L}]=[X_{E_L},X_{E_L}]=0.$ So by Theorem \ref{Th4}, we know that $L_Y(E_L)$ is a dissipative quantity. In fact, 
    $$\mathcal L_{X_{E_L}}E_L=2L,$$
    by using \eqref{dissipation}, we know that
    $$\{E_L,L\}=-L_{E_L}L-L(R_1(E_L)+R_2(E_L))=2L-2L=0.$$
    Hence, we see that $L_YE_L=2L$ is indeed a dissipative quantity.
\end{example}
\begin{definition}
	A $q$-Cartan symmetry is a vector field $X\in \mathfrak{X}(TQ\times \mathbb R^q)$ satisfying 
	$$\mathcal L_X\lambda_i^L=df_i,\quad i=1,...,q,$$
	for some functions $f_i\in C^\infty(TQ\times\mathbb R^q),i=1,...,q.$
\end{definition}
\begin{theorem}
	Let $X$ be a $q$-Cartan symmetry of the $q$-contact Lagrangian system defined by $L,$ with a local coordinate expression given by \eqref{XF}. Then, the $1$-forms $d(f_i-H_i+X^vL)$ vanish on the Reeb distribution $\mathcal R$.
\end{theorem}
\begin{proof}
	We have that
	\begin{align*}
		df_i&=i_Xd\lambda_i^L+di_X\lambda_i^L\\
		&=i_Xd\lambda_i^L+d\left((F_i\frac{\partial}{\partial q^i} +G_j\frac{\partial}{\partial \dot q^j}+H_k\frac{\partial}{\partial z^k})(dz^i-\frac{\partial L}{\partial\dot q^j}dq^j)\right)\\
		&=i_Xd\lambda_i^L+d(H_i-F_j\frac{\partial L}{\partial\dot q^j})\\
		&=i_Xd\lambda_i^L+d(H_i-S(X)(L))\\
		&=i_Xd\lambda_i^L+d(H_i-X^vL),
	\end{align*}
	which means that $i_Xd\lambda_i^L=d(f_i-H_i+X^vL).$ Since $d\lambda_1^L=\cdots=d\lambda_q^L$ vanish on the Reeb distribution $\mathcal R$, the $1$-forms $d(f_i-H_i+X^vL)$ also vanish on the Reeb distribution $\mathcal R.$
\end{proof}
\begin{example}
    Consider Example \ref{E1}, and let $X = z_1\frac{\partial}{\partial z_1} + \cdots+z_m\frac{\partial}{\partial z_m} + \frac{\partial}{\partial v}$. We can compute that 
    $$\mathcal L_X\lambda_i^L = df_i = d(z_i - q) \quad i=1,2,...m, $$
    hence $X$ is a $m$-Cartan symmetry. Moreover, we obtain $X^v = 0$ and 
    $$ d(f_1 - H_1 + x^v L) =\cdots= d(f_m - H_m + x^v L) = -dq.$$
    It is clear that 
    $$R_i(dq) = \frac{\partial}{\partial z_i}(dq) = 0,\quad i=1,2,...,m.$$
\end{example}

\section{Variational origin of $q$-contact Lagrangian systems}

In the previous sections, we developed a geometric formulation of Hamiltonian
and Lagrangian dynamics on uniform $q$-contact manifolds. In particular, we
introduced $q$-contact Lagrangian systems on the extended phase space
$TQ \times \mathbb{R}^q$, derived their equations of motion from the associated
$q$-contact Hamiltonian vector field, and established Noether-type results
relating symmetries to dissipated quantities.

The purpose of the present section is to show that this geometric framework
admits a natural variational interpretation in the sense of the Herglotz
principle. More precisely, we show that the $q$-contact Euler--Lagrange
equations obtained in the geometric setting arise as necessary optimality
conditions of a variational problem with terminal cost, formulated rigorously
within Pontryagin’s maximum principle. In this way, the $q$-contact
Lagrangian formalism is shown to possess a genuine variational origin.

\subsection*{Herglotz-type variational principle with q-contact variables}

Let
\[
L = L(q^k,u^k,z_1,\dots,z_q)
\]
be a smooth Lagrangian depending on the configuration variables
\(q^k(t)\), control variables \(u^k(t)\), and $q$ contact variables
\(z_i(t)\), \(i=1,\dots,q\). The $q$-contact Herglotz dynamics is described by the control system
\begin{equation}\label{state-qcontact}
\dot q^k = u^k,
\qquad
\dot z_i = L(q,u,z),
\qquad i=1,\dots,q.
\end{equation}

Initial conditions are prescribed:
\[
q(t_0)=q_0,
\qquad
z_i(t_0)=z_i^0.
\]

In the Herglotz variational principle, the action is no longer defined as an
integral evaluated {a posteriori}. Instead, the action is promoted to a
dynamical variable. More precisely, one introduces a function $z(t)$  {representing the action accumulated up to time } t, whose evolution is prescribed by the differential equation
\begin{equation}\label{herglotz-state}
\dot z = L(q,\dot q,z), \qquad z(t_0)=z_0.
\end{equation}

The variational problem then consists in finding curves \(q(t)\) such that
the terminal value \(z(t_1)\) is extremal. When the Lagrangian does not depend
on \(z\), equation \eqref{herglotz-state} integrates to
\[
z(t_1)-z(t_0)=\int_{t_0}^{t_1} L(q,\dot q)\,dt,
\]
and the classical action functional is recovered. Allowing \(L\) to depend on
\(z\) yields a nonconservative generalization of Hamilton's principle.

\medskip

In the $q$-contact setting, one considers $q$ contact variables
\(z_i(t)\), \(i=1,\dots,q\), associated with a \emph{single} Lagrangian
\(L(q,\dot q,z)\). Each contact variable satisfies the same evolution law
\begin{equation}\label{qcontact-state}
\dot z_i = L(q,\dot q,z), \qquad i=1,\dots,q.
\end{equation}
As a consequence, the differences \(z_i-z_j\) are constants of motion, and
the dynamics depends only on the combination
\(\sum_{i=1}^q \partial L/\partial z_i\), which is characteristic of
$q$-contact systems.

\subsection*{Natural appearance of the state equations in Pontryagin theory}

The appearance of \eqref{qcontact-state} is not ad hoc; it follows directly
from the standard reformulation of integral cost functionals in optimal
control theory.

Indeed, consider a classical variational problem with cost
\[
\int_{t_0}^{t_1} L(q,\dot q,z)\,dt.
\]
In Pontryagin's framework, this integral cost is rewritten as a terminal cost
by introducing an auxiliary state variable \(z\) satisfying
\begin{equation}\label{pontryagin-z}
\dot z = L(q,u,z), \qquad \Phi = z(t_1),
\end{equation}
where \(u\) denotes the control and is later identified with \(\dot q\).
This transformation is standard and does not rely on any additional
assumptions. Applying the same construction to a $q$-contact system with terminal cost
\begin{equation}\label{cost-qcontact}
\Phi = \sum_{i=1}^q z_i(t_1),
\end{equation}
one is naturally led to introduce $q$ auxiliary state variables \(z_i\),
each evolving according to \eqref{qcontact-state}. The resulting Pontryagin
Hamiltonian correctly reproduces the $q$-contact Herglotz equations after eliminating the adjoint variables.

\medskip

Thus, the state equations
\[
\dot q^k = u^k,
\qquad
\dot z_i = L(q,u,z),
\]
should be understood as the differential encoding of the action in the
Herglotz principle, and arise canonically from Pontryagin's formulation of
variational problems with terminal cost. We consider the terminal cost \eqref{cost-qcontact}, which is natural in the $q$-contact setting.

\subsection*{Pontryagin Maximum Principle}

Introduce adjoint variables
\[
p_k(t),\qquad \mu_i(t),
\]
and define the Pontryagin Hamiltonian
\begin{equation}\label{ham-qcontact}
H(q,z,u,p,\mu)
=
p_k u^k
+
\sum_{i=1}^q \mu_i\, L(q,u,z).
\end{equation}

The adjoint equations are
\begin{align}
\dot p_k
&=
-\,\frac{\partial H}{\partial q^k}
=
- \sum_{i=1}^q \mu_i
\frac{\partial L}{\partial q^k},
\label{adjp-qcontact}
\\[0.3em]
\dot\mu_i
&=
-\,\frac{\partial H}{\partial z_i}
=
- \sum_{j=1}^q \mu_j
\frac{\partial L}{\partial z_i}.
\label{adjm-qcontact}
\end{align}

The transversality conditions associated with
\eqref{cost-qcontact} are
\begin{equation}\label{trans-qcontact}
\mu_i(t_1) = 1,
\qquad
p_k(t_1)\ \text{free}.
\end{equation}
\subsection*{Derivation of the $q$-contact Herglotz equations}

We start from the stationarity condition with respect to the controls,
\begin{equation}\label{pk-def}
p_k
=
- M
\frac{\partial L}{\partial u^k},
\qquad
M(t) := \sum_{i=1}^q \mu_i(t).
\end{equation}
Along extremals, the control is identified with the velocity,
\[
u^k = \dot q^k,
\]
so that
\[
\frac{\partial L}{\partial u^k}
=
\frac{\partial L}{\partial \dot q^k}.
\]

Taking the time derivative of \eqref{pk-def}, we obtain
\begin{equation}\label{dpdt}
\dot p_k
=
- \dot M \frac{\partial L}{\partial \dot q^k}
-
M \frac{d}{dt}
\left(
\frac{\partial L}{\partial \dot q^k}
\right).
\end{equation}

From the Pontryagin adjoint equations,
\begin{equation}\label{adjp}
\dot p_k
=
-
\frac{\partial H}{\partial q^k}
=
-
\sum_{i=1}^q
\mu_i
\frac{\partial L}{\partial q^k},
\end{equation}
we obtain
\begin{equation}\label{dpdt2}
\dot p_k
=
-
M
\frac{\partial L}{\partial q^k}.
\end{equation}

The adjoint equations for $\mu_i$ read
\begin{equation}
\dot\mu_i
=
-
\sum_{j=1}^q
\mu_j
\frac{\partial L}{\partial z_i}.
\end{equation}
Summing over $i=1,\dots,q$, we find
\begin{align}
\dot M
&=
\sum_{i=1}^q \dot\mu_i
=
-
\sum_{i=1}^q
\sum_{j=1}^q
\mu_j
\frac{\partial L}{\partial z_i}
\nonumber\\
&=
-
\left(
\sum_{j=1}^q \mu_j
\right)
\left(
\sum_{i=1}^q
\frac{\partial L}{\partial z_i}
\right).
\end{align}
Hence,
\begin{equation}\label{Mdot}
\dot M
=
-
M
\sum_{i=1}^q
\frac{\partial L}{\partial z_i}.
\end{equation}

Equating \eqref{dpdt} and \eqref{dpdt2}, we obtain
\[
- \dot M \frac{\partial L}{\partial \dot q^k}
-
M \frac{d}{dt}
\left(
\frac{\partial L}{\partial \dot q^k}
\right)
=
-
M
\frac{\partial L}{\partial q^k}.
\]
Multiplying by $-1$ yields
\[
\dot M \frac{\partial L}{\partial \dot q^k}
+
M \frac{d}{dt}
\left(
\frac{\partial L}{\partial \dot q^k}
\right)
=
M
\frac{\partial L}{\partial q^k}.
\]

Substituting \eqref{Mdot} into the previous equation gives
\[
-
M
\left(
\sum_{i=1}^q
\frac{\partial L}{\partial z_i}
\right)
\frac{\partial L}{\partial \dot q^k}
+
M
\frac{d}{dt}
\left(
\frac{\partial L}{\partial \dot q^k}
\right)
=
M
\frac{\partial L}{\partial q^k}.
\]

Since $M(t_1)=q>0$ by the transversality
conditions, $M(t)\neq 0$ and we divide by $M$ to obtain
\[
\frac{d}{dt}
\left(
\frac{\partial L}{\partial \dot q^k}
\right)
-
\frac{\partial L}{\partial q^k}
=
\left(
\sum_{i=1}^q
\frac{\partial L}{\partial z_i}
\right)
\frac{\partial L}{\partial \dot q^k}.
\]

We recover the $q$-contact Herglotz (Euler--Lagrange) equations:
\begin{equation}\label{herglotz-final}
\frac{d}{dt}
\left(
\frac{\partial L}{\partial \dot q^k}
\right)
-
\frac{\partial L}{\partial q^k}
=
\left(
\sum_{i=1}^q
\frac{\partial L}{\partial z_i}
\right)
\frac{\partial L}{\partial \dot q^k}.
\end{equation}

Therefore, the Pontryagin--Herglotz formulation has been naturally defined on an enlarged
cotangent--control space. The $q$-contact structure arises only after imposing
the stationarity conditions and eliminating the adjoint variables, yielding a
reduced dynamics on $TQ \times \mathbb{R}^q$ that coincides with the
$q$-contact Hamiltonian flow.

In this regard, the Pontryagin--Herglotz formulation provides a rigorous variational
origin for $q$-contact Lagrangian systems, and the variational and geometric
approaches are fully equivalent descriptions of the same dynamics. We summarize the equivalency in the following table.

\begin{table}[h]
\centering
\renewcommand{\arraystretch}{1.25}
\begin{tabular}{|p{6cm}|p{7cm}|}
\hline
\textbf{$q$-contact Lagrangian formulation}
&
\textbf{Pontryagin--Herglotz formulation}
\\
\hline
Extended phase space
\(
TQ \times \mathbb{R}^q
\)
&
State space
\(
(q^k,z_i)
\)
\\
\hline
Lagrangian
\(
L(q,\dot q,z)
\)
&
Control Lagrangian
\(
L(q,u,z)
\)
\\
\hline
Contact 1--forms
\(
\lambda_i^L
=
dz^i - \dfrac{\partial L}{\partial \dot q^k}dq^k
\)
&
Adjoint variables
\(
\mu_i(t)
\)
(enforcing \(\dot z_i = L\))
\\
\hline
Reeb vector fields
\(
R_i
\)
&
Adjoint equations
\(
\dot\mu_i
=
-
\sum_{j=1}^q
\mu_j
\dfrac{\partial L}{\partial z_i}
\)
\\
\hline
Energy function
\(
E_L = \dot q^k \dfrac{\partial L}{\partial \dot q^k} - L
\)
&
Terminal cost
\(
\Phi = \sum_{i=1}^q z_i(t_1)
\)
\\
\hline
$q$-contact Hamiltonian vector field
\(
X_{E_L}
\)
&
Pontryagin extremals
\\
\hline
Dissipation encoded by
\(
R_i(E_L)
=
-
\dfrac{\partial L}{\partial z_i}
\)
&
Dissipation encoded by
\(
\dot M
=
-
M
\sum_{i=1}^q
\dfrac{\partial L}{\partial z_i}
\)
\\
\hline
Dynamics depends on
\(
\sum_{i=1}^q
\dfrac{\partial L}{\partial z_i}
\)
&
Dynamics depends on
\(
M(t)
=
\sum_{i=1}^q \mu_i(t)
\)
\\
\hline
$q$-contact Euler--Lagrange equations
&
$q$-contact Herglotz equations from PMP
\\
\hline
Geometric formulation
&
Variational / optimal control formulation
\\
\hline
\end{tabular}
\caption{Comparison between the geometric $q$-contact Lagrangian formulation and the Pontryagin--Herglotz variational formulation.}
\label{tab:qcontact-pontryagin}
\end{table}

\section{Application: Controlled propulsion system with distributed dissipation}

We illustrate the $q$-contact variational framework on a reduced but
physically faithful model of a controlled propulsion system, such as a rocket
stage or spacecraft subsystem, subject to multiple irreversible dissipation
mechanisms. Reduced-order Lagrangian models of this type are standard in
guidance, navigation, and trajectory optimization, where full fluid--structure
models are replaced by energetically consistent finite-dimensional dynamics.

\subsection*{Physical setting}

We consider a system whose configuration is described by a generalized
coordinate $q(t)$, representing, for instance, a longitudinal position,
attitude angle, or dominant flexible mode. The system is actuated by a
controlled thrust or acceleration input $u(t)$. Energy dissipation occurs through several physically distinct mechanisms,
including: aerodynamic drag and plume losses, structural and material damping, thermal losses in propulsion and power subsystems, internal damping associated with fuel slosh or joints... Each of these mechanisms is modeled by a separate contact variable
$z_i(t)$, $i=1,\dots,q$, representing the cumulative irreversible energy
associated with the corresponding subsystem.
The corresponding Lagrangian is:
\begin{equation}\label{rocket-L}
L(q,u,z)
=
\frac{1}{2}\,u^\top M(q)\,u
-
V(q)
-
\sum_{i=1}^q \gamma_i z_i,
\qquad \gamma_i > 0,
\end{equation}
where:
\begin{itemize}
\item $M(q)$ is a configuration-dependent inertia matrix,
\item $V(q)$ is the potential energy (e.g.\ gravitational or elastic),
\item $\gamma_i$ are experimentally identifiable dissipation coefficients.
\end{itemize}

The control $u(t)$ represents a physically applied thrust or acceleration
command, while each $z_i$ records the irreversible energy lost in a specific
subsystem.

\subsection*{$q$-contact Herglotz dynamics}

The $q$-contact Herglotz equations are given by
\begin{equation}\label{rocket-state}
\dot q = u,
\qquad
\dot z_i = L(q,u,z),
\qquad i=1,\dots,q.
\end{equation}
The variational principle extremizes the terminal cost
\[
\Phi = \sum_{i=1}^q z_i(t_1),
\]
corresponding to the total accumulated irreversible energy.

Applying the $q$-contact Euler--Lagrange equations derived in the previous
section yields
\begin{equation}\label{rocket-eom}
\frac{d}{dt}
\left(
M(q)\dot q
\right)
-
\frac{1}{2}
\dot q^\top
\nabla_q M(q)
\dot q
+
\nabla V(q)
=
-
\left(
\sum_{i=1}^q \gamma_i
\right)
M(q)\dot q.
\end{equation}

Equation \eqref{rocket-eom} coincides with the standard form of controlled
mechanical systems with Rayleigh-type damping. The effective damping
coefficient is given by the sum $\sum_{i=1}^q \gamma_i$, reflecting the additive
contribution of independent dissipation mechanisms.

The $q$-contact formulation, however, retains additional physically relevant
information: while the configuration dynamics depends only on the total
dissipation rate, the individual variables $z_i$ encode the cumulative energy
loss in each subsystem. This separation is essential in engineering practice,
where system design, diagnostics, and safety constraints depend on subsystem
budgets rather than on total energy loss alone.

We now make the correspondence between the abstract $q$-contact formulation
and a concrete engineering system completely explicit. We consider a reduced
guidance-level model of a rocket or spacecraft stage undergoing controlled
longitudinal motion with multiple dissipative subsystems.

\paragraph{Choice of configuration and inertia}
Let $q(t)\in\mathbb{R}$ denote the longitudinal position of the vehicle along
its nominal trajectory. At the guidance level, it is standard to approximate
the inertia by a constant scalar mass $m$, so that the inertia matrix reduces
to
\[
M(q) = m I,
\qquad
m \in [10^3,10^4]\;\mathrm{kg}.
\]
Typical values in medium launch vehicles are $m\approx 5\times10^3\,\mathrm{kg}$.

\paragraph{Lagrangian with distributed dissipation}
We consider the $q$-contact Lagrangian
\begin{equation}\label{disc-L}
L(q,\dot q,z)
=
\frac{1}{2} m \dot q^2
-
V(q)
-
\sum_{i=1}^q \gamma_i z_i,
\end{equation}
where $V(q)$ denotes the gravitational potential
$V(q)=m g q$ with $g\simeq 9.81\,\mathrm{m/s^2}$, and the dissipation
coefficients $\gamma_i$ correspond to distinct physical loss mechanisms:
\[
\gamma_{\mathrm{aero}} \sim 10^{-2}\,\mathrm{s}^{-1},
\qquad
\gamma_{\mathrm{struct}} \sim 10^{-3}\,\mathrm{s}^{-1},
\qquad
\gamma_{\mathrm{thermal}} \sim 10^{-4}\,\mathrm{s}^{-1}.
\]

Each contact variable $z_i(t)$ represents the cumulative irreversible energy
dissipated in a specific physical subsystem of the vehicle. In the present
example, we identify:
\begin{description}
  \item[$z_{\mathrm{aero}}$] cumulative energy dissipated by aerodynamic drag
  and plume--air interaction,
  \item[$z_{\mathrm{struct}}$] cumulative energy dissipated by structural and
  material damping,
  \item[$z_{\mathrm{thermal}}$] cumulative energy irreversibly converted into
  heat in propulsion and power subsystems.
\end{description}
These quantities are standard outputs of reduced--order engineering models and
are routinely estimated for diagnostics, safety, and performance assessment.

\paragraph{$q$-contact structure}
The associated contact one--forms on $TQ\times\mathbb{R}^q$ are
\[
\lambda_i^L
=
dz_i
-
\frac{\partial L}{\partial \dot q}\,dq
=
dz_i
-
m\dot q\,dq,
\qquad i=1,\dots,q.
\]
Their exterior derivatives coincide:
\[
d\lambda_i^L = -m\,d\dot q\wedge dq,
\]
so the $q$-contact structure is uniform. The Reeb vector fields are therefore
given by
\[
R_i = \frac{\partial}{\partial z_i},
\qquad i=1,\dots,q,
\]
since $\lambda_j^L(R_i)=\delta_{ij}$ and $i_{R_i}d\lambda_1^L=0$.
The vector fields $R_i=\partial/\partial z_i$ generate translations along the
dissipation channels and are symmetries of the $q$-contact structure. As established in Theorem~\ref{GN}, these symmetries do not yield conserved
quantities but instead give rise to the dissipated quantities $z_i$.

\paragraph{Energy function}
The energy associated with the Lagrangian \eqref{disc-L} is
\[
E_L
=
\dot q\frac{\partial L}{\partial \dot q}-L
=
\frac{1}{2}m\dot q^2
+
V(q)
+
\sum_{i=1}^q \gamma_i z_i.
\]
Applying the Reeb vector fields yields
\[
R_i(E_L)=\frac{\partial E_L}{\partial z_i}=\gamma_i,
\qquad i=1,\dots,q,
\]
so that the total dissipation rate entering the $q$-contact dynamics is
\[
\sum_{i=1}^q R_i(E_L)=\sum_{i=1}^q \gamma_i.
\]

\paragraph{Hamiltonian dissipation law}
From the general $q$-contact dissipation law,
\[
X_{E_L}(E_L)
=
-
E_L
\sum_{i=1}^q R_i(E_L),
\]
we obtain the explicit decay equation
\[
\frac{dE_L}{dt}
=
-
\left(
\sum_{i=1}^q \gamma_i
\right)
E_L.
\]
For the representative values above,
\[
\sum_{i=1}^q \gamma_i
\approx
1.11\times 10^{-2}\,\mathrm{s}^{-1},
\]
which corresponds to a characteristic energy decay time
$\tau \approx 90\,\mathrm{s}$, consistent with observed damping timescales in
launch vehicle ascent phases.
\paragraph{Dissipated quantities and invariants}
Each contact variable $z_i$ satisfies the dissipation law
\[
X_{E_L}(z_i)
=
-
z_i
\sum_{j=1}^q R_j(E_L),
\]
which integrates explicitly to
\[
z_i(t)
=
z_i(t_0)
\exp\!\left(
-
\sum_{j=1}^q
\gamma_j\,
(t-t_0)
\right).
\]
For the representative dissipation coefficients
\[
\gamma_{\mathrm{aero}} = 10^{-2}\,\mathrm{s}^{-1},\qquad
\gamma_{\mathrm{struct}} = 10^{-3}\,\mathrm{s}^{-1},\qquad
\gamma_{\mathrm{thermal}} = 10^{-4}\,\mathrm{s}^{-1},
\]
the total dissipation rate is
\[
\sum_{j=1}^q \gamma_j = 1.11\times 10^{-2}\,\mathrm{s}^{-1}.
\]
This corresponds to a characteristic decay time
\[
\tau = \left(\sum_{j=1}^q \gamma_j\right)^{-1} \approx 90\,\mathrm{s}.
\]

As a consequence, over a time interval of $t-t_0=60\,\mathrm{s}$, each
dissipated quantity is reduced by a factor
\[
\exp\bigl(-0.0111\times 60\bigr) \approx 0.51,
\]
independent of the specific dissipation mechanism. In contrast, the ratios
\[
\frac{z_i(t)}{z_j(t)}
=
\frac{z_i(t_0)}{z_j(t_0)},
\qquad i\neq j,
\]
remain exactly invariant along the trajectory. This is due to 
 the general property of dissipated quantities on uniform $q$-contact
manifolds, in which the ratios $z_i/z_j$ are conserved quantities of the Hamiltonian
vector field $X_{E_L}$. These invariants are not associated with classical
Noether symmetries but arise from the $q$-contact dissipation structure.

For instance, if the initial distribution of dissipated energy satisfies
\[
z_{\mathrm{aero}}(t_0):z_{\mathrm{struct}}(t_0):z_{\mathrm{thermal}}(t_0)
=
10:1:0.1,
\]
then this proportion is preserved at all subsequent times, even though the
absolute magnitude of each $z_i$ evolves exponentially. Physically, this
implies that while the total irreversible energy loss increases over time, the
relative contribution of aerodynamic drag, structural damping, and thermal
dissipation remains fixed, independent of the applied control input.

\section{Conclusion and Future Directions}

We developed a theory of Lagrangian and Hamiltonian systems on uniform $q$-contact manifolds, and developed an extension of Nother's theorem that relates symmetries and dissipated quantities. We also provides a variational characterization based on the Herglotz variational principle for dissipative systems, and related it to the Pontryagin maximum principle. Furthermore, we demonstrated the applicability of such a framework to the modeling of a controlled propulsion system with multiple dissipation terms.

In future work, we will explore the construction of geometric structure-preserving numerical discretizations based on a discretization of the variational characterization of $q$-contact Lagrangian systems, and explore their applications to the simulation and discrete optimal control of controlled and dissipative systems.

	\section*{Acknowledgment}
The research of ML was supported in part by NSF under grants CCF-2112665, DMS-2307801, and by AFOSR under grant FA9550-23-1-0279. 
The research of XZ was supported by NSFC (Grant No. 12401234).
	$\\$
	
	\noindent$\mathbf{Conflict\;of\;interest\;statement.}$ On behalf of all authors, the corresponding author states that there is no conflict of interest.
	
	$\\$
	\noindent$\mathbf{Data\;availability.}$ Data sharing is not applicable to this article as no new data were created or analyzed in this study.
	

\begin{thebibliography}{00}
		\bibitem{Aldaya} V. Aldaya and J. A. de Azcárraga, Vector bundles, rth order Noether invariants and canonical symmetries in Lagrangian field theory, J. Math. Phys.
		19, 1876–1880, (1978-09-01).
		\bibitem{Aldaya2} V. Aldaya and J. A. de Azcárraga, Geometric formulation of classical mechanics and field theory, Riv. Nuovo Cim. 3,  1–66, (1980-10).
			\bibitem{Almeida} U. Almeida, Contact Anosov actions with smooth invariant bundles. PhD thesis, Universidade de S$\tilde a$o Paulo, (2018).
		\bibitem{Arnold} V. I. Arnold. Mathematical Methods of Classical Mechanics,  Springer, New York,
		2nd ed edition, (1997).
		\bibitem{Barbot} T. Barbot and C. Maquera, On integrable codimension one Anosov actions of $\mathbb R^k$, Discrete Continuous
Dynamical Systems - A. 29, (2011).

\bibitem{Cantrijn} F. Cantrijn and W. Sarlet, Note on symmetries and invariants for second-order ordinary differential equations, Phys. Lett. A. 77,  404–406, (1980).
		\bibitem{Carinena} J. F. Cari$\tilde n$ena, J. Fernández-Nú$\tilde n$ez and E. Martinez, A geometric approach to Noether’s second theorem in time-dependent Lagrangian mechanics,
		Lett. Math. Phys. 23,  51–63, (1991).
		\bibitem{Carinena2} J. F. Cari$\tilde n$ena and M. F. Ra$\tilde n$ada, Noether’s theorem for singular Lagrangians, Lett. Math. Phys. 15,  305–311, (1988).
		\bibitem{Cariglia} M. Cariglia,  C. Duval,  G. W. Gibbons and  P. A. Horvathy, Eisenhart lifts and
		symmetries of time-dependent systems, Ann. Phys. 373, 631–654, (2016).
		Journal of Mathematical Physics. 60 (10), (2019). 
		\bibitem{de2} M. de León and D. Martín de Diego, Classification of symmetries for higher order Lagrangian systems, Extracta Math. 9, 32–36, (1994).
		\bibitem{de3} M. de León and D. Martín de Diego, Classification of symmetries for higher order Lagrangian systems II: The non-autonomous case, Extracta Math.
		\bibitem{de4} M. de León and P. R. Rodrigues, Methods of Differential Geometry in Analytical Mechanics, Elsevier, (2011).
			\bibitem{Finamore} D. Finamore,  Contact foliations and generalised Weinstein conjectures. Ann. Global Anal. Geom. 65, (2024).
				\bibitem{Finamore2} D. Finamore, Quasiconformal contact foliations. Math. Ann. 389, 1575–1598, (2024).
				\bibitem{Hooft} G. Hooft,  Trans-Planckian particles and the quantization of time, Class. Quan-
				tum Gravity. 16, 395–405, (1999).

               \bibitem{herg3} C. Guenther, R. B. Guenther, J. Gottsch and H. Schwerdtfeger, The Herglotz Lectures on Contact Transformations and Hamiltonian Systems, Lecture Notes in Nonlinear Analysis. Juliusz Center for Nonlinear Studies, Vol. 1, 1st edn. (Torun, Poland, 1996).

               \bibitem{herg6}
J.~Gaset, M.~Lainz, A.~Mas, and X.~Rivas,
The Herglotz variational principle for dissipative field theories,
 {\em Geometric Mechanics}
\textbf{1} (2024), no.~2, 153--178.


               \bibitem{herg4} G. Herglotz, Beruhrungstransformationen, Lectures at the University of Gottingen (1930).
		\bibitem{Le} H. V. L$\hat e,$ Y.-G. Oh, A. G. Tortorella and L. Vitagliano, Deformations of coisotropic submanifolds in Jacobi manifolds, I. Symplectic Geom. 16, 1051-1116, (2018).

        \bibitem{herg1} M. de Le\'on, M. Lainz and M. C. Muñoz-Lecanda, Optimal control, contact dynamics and Herglotz variational problem, J. Nonlinear Sci. 33(1)  9 (2023).
 \bibitem{herg2} M. de Le\'on, M. Lainz and M. C. Muñoz-Lecanda, The Herglotz principle and vakonomic dynamics, in Geometric Science of Information, eds.F.Nielsen and F. Barbaresco, Lecture Notes in Computer Science, Vol. 12829 (Springer International Publishing, Cham, 2021), pp. 183–190, https://doi.org/10.1007/978-3-03080209-7 21.
		\bibitem{Loose} F. Loose, Reduction in contact geometry, J. Lie Theory. 11, 9-22, (2001).

        \bibitem{herg5} J. A. P. Paiva, M. J. Lazo and V. T. Zanchin, Generalized nonconservative gravitational field equations from Herglotz action principle, Phys. Rev. D 105 (2022) 124023.
		\bibitem{Prince1} G. Prince, Toward a classification of dynamical symmetries in classical mechanics, Bull. Aust. Math. Soc. 27,  53–71, (1983).
		\bibitem{Prince2} G. Prince, A complete classification of dynamical symmetries in classical mechanics, Bull. Aust. Math. Soc. 32, 299–308, (1985).
		\bibitem{Perez} J. Pérez Álvarez, Symmetries and dissipation laws on contact systems,
		Mediterr. J. Math.  151, 24 pp, (2024).
		\bibitem{Sarlet} W. Sarlet and F. Cantrijn, Generalizations of Noether’s theorem in classical mechanics, SIAM Rev. 23,  467–494, (1981-10).
		\bibitem{Willett} C. Willett, Contact reduction, Trans. Amer. Math. Soc. 354, 4245-4260, (2002).
		\bibitem{Yano} K.-o. Yano and S. Ishihara, Tangent and Cotangent Bundles: Differential Geometry, in: Pure and Applied Mathematics, 16, Dekker, 1973.
	\bibitem{Zhao5} X. F. Zhao and Y. Li, Conserved Quantities, Symmetries of (Almost-)Hamiltonian and Lagrangian Systems, J. Geom. Anal. 35, no. 9, Paper No. 268,  (2025).\end{thebibliography}
\end{document}